\documentstyle[preprint,aps]{revtex}
\tightenlines
\begin{document}
\title{Modeling 1/f noise }
\author{B. Kaulakys$^{1,2}$ and T. Me\v skauskas$^{1,3}$}
\address{$^1$Institute of Theoretical Physics and Astronomy, A. Go\v stauto 12, 2600
Vilnius, Lithuania \\
$^2$Department of Physics, Vilnius University, Saul\.etekio al. 9, 2040
Vilnius, Lithuania \\
$^3$Department of Mathematics, Vilnius University, Naugarduko 24, 2006
Vilnius, Lithuania \\
}
\date{Received 29 June 1998}
\maketitle

\begin{abstract}

{\bf Physical Review E Vol. 58, No. 6, p. 7013-19 }\vspace{1cm}

The noise of signals or currents consisting from a sequence of pulses,
elementary events or moving discrete objects (particles) is analyzed. A
simple analytically solvable model is investigated in detail both
analytically and numerically. It is shown that 1/f noise may result from
the statistics of the pulses transit times with random increments of the time
intervals between the pulses.
The model also serves as a basis for revealing parameter
dependences of 1/f noise and allows one to make some generalizations. As a
result the intensity of 1/f noise is expressed through the distribution and
characteristic functions of the time intervals between the subsequent
transit times of the pulses. The conclusion that 1/f noise may result from
the clustering of the signal pulses, elementary events or particles can be
drawn from the analysis of the model systems.

\noindent PACS number(s): 05.40.+j, 02.50.-r, 72.70.+m
\end{abstract}

\pacs{05.40.+j, 02.50.-r, 72.70.+m}

\section{\bf Introduction }

The omnipresence of 1/f noise is one of the oldest puzzle of the
contemporary physics. During more than 70 years since the first observation
by Johnson, long-memory processes with long-term correlations have been
observed in many types of systems from physics, technology, biology,
astrophysics, geophysics and sociology (see \cite
{hooge81,musha76,press78,koba82} and references herein). Recently 1/f noise
was discovered in human cognition \cite{gilden95}, human coordination \cite
{chen97} and even in distribution of prime numbers \cite{wolf97}.

Despite of the widespread occurrence of fluctuations of signals and
variables exhibiting $1/f^\delta $ ($\delta \simeq 1$) behavior of the power
spectral density $S\left( f\right) $ at low frequencies in large diversity
of systems no generally recognized explanation of the ubiquity of 1/f noise
is still proposed. Physical models of 1/f noise in some physical systems are
usually very specialized, complicated (see \cite
{hooge81,musha76,press78,koba82} and references herein) and they do not
explain the omnipresence of the processes with $1/f^\delta $ spectrum \cite
{icnf95,icnf97,upon97}. Note also some mathematical analysis \cite
{mandelbr68}, models and algorithms of generation of the processes with 1/f
noise \cite{jensen91,kumic94,sinha96}. These models also expose some
shortcomings: they are very specific, formal (like ''fractional Brownian
motion'' or half-integral of a white noise signal) or unphysical. They can
not, as a rule, be solved analytically and they do not reveal the origin as
well as the necessary and sufficient conditions for the appearance of 1/f
type fluctuations.

In such a situation the simple analytically solvable model system generating
1/f noise may essentially influence in the reveal of the origin and essence
of the effect. Here we present a model which generates 1/f noise in any
desirably wide range of frequency. Our model is a result of the search of
necessary and sufficient conditions for the appearance of 1/f fluctuations
in simple systems affected by the random external perturbations initiated in
\cite{kaul97} and originated from the observation of the transition from
chaotic to nonchaotic behavior in the ensemble of randomly driven systems
\cite{kaul95}. Contrary to the McWhorter model \cite{surd39} based on the
superposition of large number of Lorentzian spectra and requiring a very
wide distribution of relaxation times, our model contains only one
relaxation rate $\gamma $ and can have an exact 1/f spectrum in any
desirably wide range of frequency. The model may be used as a basis for the
checking of assumptions made in the derivation of 1/f noise spectrum for
different systems. Furthermore, it allows us to make a heuristic presumption
for the generalizations of the theory of 1/f noise. Numerical simulations
and comparisons with analytical results confirm this supposition.

\section{\bf The model }

In many cases, the intensity of some signal or current can be represented by
a sequence of random (however, as a rule, mutually correlated) pulses or
elementary events $A_k\left( t-t_k\right) $. Here the function $A_k\left(
\varphi \right) $ represents the shape of the $k$ pulse having an influence
to the signal $I\left( t\right) $ in the region of transit time $t_k$. The
signal or intensity of the current of particles in some space cross-section
may, therefore, be expressed as
$$
I\left( t\right) =\sum\limits_kA_k\left( t-t_k\right) .\eqno{(1)}
$$
It is easy to show that the shapes of the pulses influent mainly on the high
frequency, $f\geq \Delta t_p$ with $\Delta t_p$ being the characteristic
pulse length, power spectral density while fluctuations of the pulse
amplitudes result, as a rule, in the white or Lorentzian but not 1/f noise
\cite{lukes61}. Therefore, we restrict our analysis to the noise due to the
correlations between the transit times $t_k$. In such an approach we can
replace the function $A_k\left( t-t_k\right) $ by the Dirac delta function $
\delta \left( t-t_k\right) $ and the signal express as
$$
I\left( t\right) =\sum\limits_k\delta \left( t-t_k\right) .\eqno{(2)}
$$
This model also corresponds to the flow of identical objects: electrons,
photons, cars and so on. On the other hand, fluctuations of the amplitudes $
A_k$ may result in the additional noise but can not reduce 1/f noise we are
looking for.

The power spectral density of the current (2) is
$$
S\left( f\right) =\lim \limits_{T\rightarrow \infty }\left\langle \frac 2T
\left| \sum\limits_{k=k_{\min }}^{k_{\max }}e^{-i2\pi ft_k}\right|
^2\right\rangle =\lim \limits_{T\rightarrow \infty }\left\langle \frac 2T
\sum\limits_k\sum\limits_{q=k_{\min }-k}^{k_{\max }-k}e^{i2\pi f\left(
t_{k+q}-t_k\right) }\right\rangle \eqno{(3)}
$$
where $T$ is the whole observation time interval, $k_{\min }$ and $k_{\max }$
are minimal and maximal values of index $k$ in the interval of observation
and the brackets $\left\langle ...\right\rangle $ denote the averaging over
realizations of the process.

In this approach the power spectral density of the signal depends on the
statistics and correlations of the transit times $t_k$ only. It is well
known that sequence of random, Poisson, transit times generates white (shot)
noise \cite{lukes61}. The sequence of transit times $t_k$ with random
increments, $t_k=t_{k-1}+\bar \tau +\sigma \varepsilon _k$ (where $\bar \tau
$ is the average time interval between pulses, $\left\{ \varepsilon
_k\right\} $ denotes the sequence of uncorrelated normally distributed
random variables with zero expectation and unit variance, i.e. the white
noise source, and $\sigma $ is the standard deviation of white noise)
results in the Lorentzian spectra \cite{kaul97}. Here we will consider
sequences of the transit times with random increments of the time intervals
between pulses, $\tau _k=\tau _{k-1}+\sigma \varepsilon _k$, where $\tau
_k=t_k-t_{k-1}$. It is natural to restrict in some way the infinite Brownian
increase or decrease of the intervals $\tau _k$, e.g. by the introduction of
the relaxation to the average period $\bar \tau $ rate $\gamma $. So, we
have the recurrent equations for the transit times
$$
\left\{
\begin{array}{ll}
t_k= & t_{k-1}+\tau _k, \\
\tau _k= & \tau _{k-1}-\gamma \left( \tau _{k-1}-\bar \tau \right) +\sigma
\varepsilon _k.
\end{array}
\right. \eqno{(4)}
$$

The simplest physical interpretation of the model (4) corresponds to one
particle moving in the closed contour with the period of the drift of the
particle round the contour fluctuating (due to the external random
perturbations) about the average value $\bar \tau $ \cite{kaul98}.

\section{\bf Solutions }

An advantage of the model (4) is that it may be solved analytically. So, an
iterative solution of Eqs. (4) results in an expression for the period
$$
\tau _k=\bar \tau +\left( \tau _0-\bar \tau \right) \left( 1-\gamma \right)
^k+\sigma \sum_{j=1}^k\left( 1-\gamma \right) ^{k-j}\varepsilon _j\eqno{(5)}
$$
where $\tau _0$ is the initial period. The dispersion of the period $\tau _k$
is
$$
\sigma _\tau ^2\left( k\right) \equiv \left\langle \tau _k^2\right\rangle
-\left\langle \tau _k\right\rangle ^2=\frac{\sigma ^2\left[ 1-\left(
1-\gamma \right) ^{2k}\right] }{2\gamma \left( 1-\gamma /2\right) }\simeq
\left\{
\begin{array}{ll}
\sigma ^2k, & 2k\gamma \ll 1 \\
\sigma ^2/2\gamma , & 2k\gamma \gg 1.
\end{array}
\right. \eqno{(6)}
$$
Therefore, after the characteristic transition to the stationary process
time, $t_{tr}=\bar \tau /\gamma $, the dispersion of the period approaches
the limiting value $\sigma _\tau ^2=\sigma ^2/2\gamma $.

After some algebra we can also obtain an explicit expression for the transit
times $t_k$ ($k\geq 1$),
$$
t_k=t_0+k\bar \tau +\frac{1-\gamma }\gamma \left[ 1-\left( 1-\gamma \right)
^k\right] \left( \tau _0-\bar \tau \right) +\frac \sigma \gamma
\sum_{l=1}^k\left[ 1-\left( 1-\gamma \right) ^{k+1-l}\right] \varepsilon _l,
\eqno{(7)}
$$
where $t_0$ is the initial time. The dispersion of the transit time $t_k$ is
$$
\sigma _t^2\left( k\right) \equiv \left\langle t_k^2\right\rangle
-\left\langle t_k\right\rangle ^2=\frac{\sigma ^2}{\gamma ^2}\left\{ k-2
\frac{1-\gamma }\gamma \left[ 1-\left( 1-\gamma \right) ^k\right] +\left(
1-\gamma \right) ^2\frac{1-\left( 1-\gamma \right) ^{2k}}{1-\left( 1-\gamma
\right) ^2}\right\}
$$

$$
=\left\{
\begin{array}{ll}
\sigma ^2\left( k/6+k^2/2+k^3/3+...\right) , & 2\gamma k\ll 1, \\
\left( \sigma /\gamma \right) ^2\left( k-3/2\gamma +5/4\pm ...\right) , &
2\gamma k\gg 1.
\end{array}
\right. \eqno{(8)}
$$
At $k\gg \gamma ^{-1}$ Eq. (7) generates the stationary time series. The
difference of the transit times $t_{k+q}$ and $t_k$ in Eq. (3) for $\tau _0=
\bar \tau $ or $2\gamma k\gg 1$ is
$$
t_{k+q}-t_k=\bar \tau q+\frac \sigma \gamma \left\{ \left[ 1-\left( 1-\gamma
\right) ^q\right] \sum_{l=1}^k\left( 1-\gamma \right) ^{k+1-l}\varepsilon
_l+\sum_{l=k+1}^{k+q}\left[ 1-\left( 1-\gamma \right) ^{k+q+1-l}\right]
\varepsilon _l\right\} ,q\geq 0.\eqno{(9)}
$$
The dispersion of this times difference equals
$$
\left\langle \left( t_{k+q}-t_k\right) ^2\right\rangle -\bar \tau ^2q^2=
\frac{\sigma ^2}2g\left( q\right) \eqno{(10)}
$$
where
$$
g\left( q\right) =\frac 2{\gamma ^2}\left\{ \left[ 1-\left( 1-\gamma \right)
^q\right] ^2\sum_{l=1}^k\left( 1-\gamma \right) ^{2l}+\sum_{l=1}^q\left[
1-\left( 1-\gamma \right) ^l\right] ^2\right\} ,\quad q\geq 0.\eqno{(11)}
$$
Summation in Eq. (11) results in
$$
g\left( q\right) =\frac 2{\gamma ^2}\left\{ q-\frac{\left( 1-\gamma \right)
\left[ 1-\left( 1-\gamma \right) ^q\right] }{1-\left( 1-\gamma \right) ^2}
\left\{ 2+\left[ 1-\left( 1-\gamma \right) ^q\right] \left( 1-\gamma \right)
^{2k+1}\right\} \right\} .\eqno{(12)}
$$
At $\gamma q\ll 1$
$$
g\left( q\right) =\left\{
\begin{array}{ll}
\left( 2k+1\right) q^2+q/3+2q^3/3, & 2\gamma k\ll 1, \\
\left( \frac 1\gamma +\frac 12\right) q^2+\frac 13q-\frac 13q^3, & 2\gamma
k\gg 1,
\end{array}
\right. \eqno{(12a)}
$$
while for $2\gamma k\gg 1$ we have
$$
g\left( q\right) =\frac 2{\gamma ^2}\left[ q-2\frac{\left( 1-\gamma \right)
\left[ 1-\left( 1-\gamma \right) ^q\right] }{1-\left( 1-\gamma \right) ^2}
\right] \eqno{(13)}
$$

$$
\simeq \left\{
\begin{array}{ll}
\left( \frac 1\gamma +\frac 12\right) q^2+\frac 13q-\frac 13q^3, & \gamma
q\ll 1, \\
\frac 2{\gamma ^2}\left( q+\frac 12\right) -\frac 2{\gamma ^3}+..., & q\gg
\gamma ^{-1}\gg 1.
\end{array}
\right. \eqno{(13a)}
$$
Note that for $q<0$ we should replace in Eq. (9)--(13) $q$ by $\left|
q\right| $and $k$ by $k-\left| q\right| $. Therefore, the function $g(q)$ at
$k-\left| q\right| \gg \gamma ^{-1}$ is even, i.e. $g(-q)=g(q)$.

Substituting Eq. (9) into Eq. (3) and replacing the summations in the
exponents by the multiplications of the exponents we have the following
expression for the power spectral density of the current
$$
\begin{array}{c}
S\left( f\right) =\lim \limits_{T\rightarrow \infty }\left\langle
\frac 2T\sum\limits_k\sum\limits_{q=k_{\min }-k}^{k_{\max }-k}e^{i2\pi f\bar
\tau q}\prod\limits_{l=1}^k\exp \left\{ i\frac{2\pi f\sigma }\gamma \left[
1-\left( 1-\gamma \right) ^q\right] \left( 1-\gamma \right)
^{k+1-l}\varepsilon _l\right\} \right. \\ \left.
\prod\limits_{l=k+1}^{k+q}\exp \left\{ i\frac{2\pi f\sigma }\gamma \left[
1-\left( 1-\gamma \right) ^{k+q+1-l}\right] \varepsilon _l\right\}
\right\rangle .
\end{array}
\eqno{(3a)}
$$

The average over realizations of the process coincides with the average over
the distribution of the random variables $\varepsilon _l$. Using the fact
that random variables $\varepsilon _l$ are independent and mutually
uncorrelated we can fulfill the averaging over every random variable $
\varepsilon _l$ independently. Therefore, Eq. (3a) may be rewritten in the
form
$$
\begin{array}{c}
S\left( f\right) =\lim \limits_{T\rightarrow \infty }
\frac 2T\sum\limits_{k,q}e^{i2\pi f\bar \tau q}\prod\limits_{l=1}^k\left
\langle \exp \left\{ i\frac{2\pi f\sigma }\gamma \left[ 1-\left( 1-\gamma
\right) ^q\right] \left( 1-\gamma \right) ^{k+1-l}\varepsilon _l\right\}
\right\rangle \\ \prod\limits_{l=k+1}^{k+q}\left\langle \exp \left\{ i\frac{
2\pi f\sigma }\gamma \left[ 1-\left( 1-\gamma \right) ^{k+q+1-l}\right]
\varepsilon _l\right\} \right\rangle .
\end{array}
\eqno{(3b)}
$$

The result of the averaging of the exponent $\exp \left\{ ic\varepsilon
_l\right\} $ (with $c$ being a constant) over the normally distributed
random variable $\varepsilon _l$ with zero expectation and unit variance is
$$
\left\langle e^{ic\varepsilon _l}\right\rangle =\int\limits_{-\infty
}^{+\infty }e^{ic\varepsilon _l}\frac 1{\sqrt{2\pi }}e^{-\varepsilon
_l^2/2}d\varepsilon _l=e^{-c^2/2}.
$$
Therefore, after the averaging over the normal distribution of the random
variables $\varepsilon _l$ Eq. (3b) takes the form
$$
\begin{array}{c}
S\left( f\right) =\lim \limits_{T\rightarrow \infty }
\frac 2T\sum\limits_{k,q}e^{i2\pi f\bar \tau q}\prod\limits_{l=1}^k\exp
\left\{ -\frac{2\pi ^2f^2\sigma ^2}{\gamma ^2}\left[ 1-\left( 1-\gamma
\right) ^q\right] ^2\left( 1-\gamma \right) ^{2\left( k+1-l\right) }\right\}
\\ \prod\limits_{l=k+1}^{k+q}\exp \left\{ -\frac{2\pi ^2f^2\sigma ^2}{\gamma
^2}\left[ 1-\left( 1-\gamma \right) ^{k+q+1-l}\right] ^2\right\} .
\end{array}
\eqno{(3c)}
$$

Transition in Eq. (3c) from the multiplications of the exponents to the
summations in the exponents and transformations in analogy with Eq. (11) of
the two sums' summation indexes $l\rightarrow k+1-l$ and $l\rightarrow
k+q+1-l$, respectively, yield according to Eq. (11) the final expression for
the power spectral density
$$
S\left( f\right) =\lim \limits_{T\rightarrow \infty }\frac 2T
\sum\limits_{k,q}e^{i2\pi f\bar \tau q-\pi ^2f^2\sigma ^2g\left( q\right) }.
\eqno{(14)}
$$

Since the expansion of the function $g\left( q\right) $ in powers of $\gamma
\left| q\right| \ll 1$ at $2\gamma k\gg 1$ according to Eqs. (12) and (13)
is
$$
g\left( q\right) =\frac 1\gamma q^2-\frac 13\left| q\right| ^3+\frac 12
q^2\pm \ldots .,\eqno{(15)}
$$
for $f\ll f_\tau =\left( 2\pi \bar \tau \right) ^{-1}$ and $f<f_2=2\sqrt{
\gamma }/\pi \sigma $ we can replace the summation in Eq. (14) by the
integration
$$
S\left( f\right) =2\bar I\int\limits_{-\infty }^{+\infty }e^{i2\pi f\bar \tau
q-\pi ^2f^2\sigma ^2g\left( q\right) }dq.\eqno{(16)}
$$
where $\bar I=\lim \limits_{T\rightarrow \infty }\left( k_{\max }-k_{\min
}+1\right) /T=\bar \tau ^{-1}$ is the averaged current.

Furthermore, at $f\gg f_1=\gamma ^{3/2}/\pi \sigma $ it is sufficient to
take into account only the first term of expansion (15), $g\left( q\right)
=q^2/\gamma $. Integration in Eq. (16) hence yields to 1/f spectrum
$$
S\left( f\right) =2\bar I\int\limits_{-\infty }^{+\infty }\exp \left[ i2\pi f
\bar \tau q-\frac{\left( \pi f\sigma \right) ^2}\gamma q^2\right] dq=\bar I^2
\frac{\alpha _H}f,\quad f_1<f<f_2,f_\tau \eqno{(17)}
$$
where $\alpha _H$ is a dimensionless constant (the Hooge parameter)
$$
\alpha _H=\frac 2{\sqrt{\pi }}Ke^{-K^2},\quad K=\frac{\bar \tau }{\sqrt{2}
\sigma _\tau }=\frac{\bar \tau \sqrt{\gamma }}\sigma .\eqno{(18)}
$$

Using an expansion of the function $g(q)$ at $\gamma q\gg 1$ according to
expression (13), $g(q)=2q/\gamma ^2$, we obtain from Eq. (16) the Lorentzian
power spectrum density for $f<f_1$
$$
S\left( f\right) =2\bar I\frac{\sigma ^2}{\bar \tau ^2\gamma ^2}\frac 1{
1+\left( \pi f\sigma ^2/\bar \tau \gamma ^2\right) ^2}=\bar I^2\frac{4\tau
_{rel}}{1+\tau _{rel}^2\omega ^2}.\eqno{(19)}
$$
Here $\omega =2\pi f$ and $\tau _{rel}=D_t=\sigma ^2/2\bar \tau \gamma ^2$
is the ''diffusion'' coefficient of the time $t_k$ according to Eqs. (7) and
(8). The model is, therefore, free from the unphysical divergence of the
spectrum at $f\rightarrow 0$; for $f\ll f_0=\bar \tau \gamma ^2/\pi \sigma
^2=1/2\pi \tau _{rel}$ we have from Eq. (19) the white noise
$$
S(f)=\bar I^2\left( 2\sigma ^2/\bar \tau \gamma ^2\right) .\eqno{(20)}
$$

Therefore, the model containing only one relaxation rate $\gamma $ for
sufficiently small parameter $\gamma $ can produce an exact 1/f-like
spectrum in any desirably wide range of frequency, $f_1<f<f_2,f_\tau $.
Furthermore, due to the contribution to the transit times $t_k$ of the large
number of the very separated in time random variables, $\varepsilon _l$ ($
l=1,2,...k$), our model represents a 'long-memory' random process.

\section{\bf Generalizations and numerical analysis }

Eqs. (16)--(20) describe quite well the power spectrum of the random process
(4). As an illustrative example in Fig. 1 the numerically calculated power
spectral density averaged over five realizations of the process (4) is
compared with the analytical calculations according to Eqs. (16)--(20). The
analytical results are in good agreement with the numerical simulations.
Note, that analytical results predict not only the slope and intensity of
1/f noise but the frequency range $f_1\div f_2,f_\tau $ of 1/f noise and
intensity of the very low frequency, $f\ll f_0$, white noise (20) as well.

This model may also be generalized for the non-Gaussian and for the
continuous perturbations of the systems' parameters resulting in the
fluctuations of the period $\tau $. So, for perturbations by the
non-Gaussian sequence of random impacts $\left\{ \varepsilon _k\right\} $
with zero expectations Eqs. (1)--(13) remain valid. Only the result (14) of
the averaging over realizations of the process in the case of the
non-Gaussian perturbations may have different form. Consider now such a
situation in more detail.

The power spectral density (3) may be rewritten in the form
$$
S\left( f\right) =2\bar I\left\langle \sum_qe^{i2\pi f\tau _k\left( q\right)
q}\right\rangle \eqno{(21)}
$$
where the transit times $t_{k+q}$ and $t_k$ difference is expressed as
$$
t_{k+q}-t_k=\sum\limits_{l=k+1}^{k+q}\tau _k=\tau _k\left( q\right) q,\quad
q\geq 0\eqno{(22)}
$$
and the brackets denote the averaging over the time (index $k$) and over the
realizations of the process. Here $\tau _k\left( q\right) \equiv \left(
t_{k+q}-t_k\right) /q$ is the averaged time interval between the subsequent
transit times in the time interval $t_k\div t_{k+q}$. Note that for the slow
(diffusive-like) fluctuations of the averaged interval $\tau _k\left(
q\right) $ with the change of the index $k$ Eq. (22) is valid also when $q<0$
, i.e. $t_{k+q}-t_k=\tau _{k-q}\left( q\right) q\simeq \tau _k\left(
q\right) q$, $q<0$. At $2\pi f\tau _k\left( q\right) \ll 1$ we may replace
the summation in Eq. (21) by the integration and do not take into account
the dependence of $\tau _k\left( q\right) $ on $q$. In such a case Eq. (21)
yields
$$
S\left( f\right) =2\bar I\left\langle \int\limits_{-\infty }^{+\infty
}e^{i2\pi f\tau _kq}dq\right\rangle =2\bar I\int\limits_{-\infty }^{+\infty
}\left\langle e^{i2\pi f\tau _kq}\right\rangle dq.\eqno{(23)}
$$
Here the averaging over $k$ and over the realizations of the process
coincides with the averaging over the distribution of the periods $\tau _k$,
i.e.
$$
\left\langle e^{i2\pi fq\tau _k}\right\rangle =\int\limits_{-\infty
}^{+\infty }e^{i2\pi fq\tau }\psi \left( \tau \right) d\tau =\chi _\tau
\left( 2\pi fq\right) \eqno{(24)}
$$
where $\psi \left( \tau \right) $ is the periods $\tau _k$ distribution
density and $\chi _\tau \left( \vartheta \right) $ is the characteristic
function of the distribution $\psi \left( \tau \right) $. Taking into
account the property of the characteristic function
$$
\int\limits_{-\infty }^{+\infty }\chi _\tau \left( \vartheta \right)
d\vartheta =2\pi \psi \left( 0\right)
$$
we have from Eqs. (23) and (24) the final expression for the power spectral
density
$$
S\left( f\right) =2\bar I\psi \left( 0\right) /f.\eqno{(25)}
$$

Substituting into Eq. (25) the value $\psi \left( 0\right) =\exp \left( -
\bar \tau ^2/2\sigma _\tau ^2\right) /\sqrt{2\pi }\sigma _\tau $ for the
Gaussian distribution of the periods $\tau _k$ we recover the result
(17)--(18).

Since different processes result in the Gaussian distribution it is likely
that perturbation by the non-Gausssian impacts $\left\{ \varepsilon
_k\right\} $ in Eq. (4) yields nevertheless the Gaussian distribution of the
periods $\tau _k$. For the demonstration of such statement and validity of
the approach (21)--(25) we have performed numerical analysis of the model
(1)--(4) for different distributions of the perturbations $\left\{
\varepsilon _k\right\} $. Figures (2) and (3) represent the calculated power
spectral densities for the rectangular (uniform) and asymmetric $\chi _3^2$
distributions of the sequence $\left\{ \varepsilon _k\right\} $ with zero
expectations and the same variances and other parameters like those for Fig.
1. We can notice only the slight dependence of the spectra on the
distribution function of the perturbing impacts $\left\{ \varepsilon
_k\right\} $ with the same expectations and variances. These results confirm
also the presumptions made in the derivation (21)--(24) of the 1/f noise
intensity (25).

\section{\bf Concluding remarks }

Analysis of the exactly solvable model of 1/f noise display main features of
the noise and serves as a basis for revealing of the origin and parameter
dependences of the flicker noise. This allows us to make generalizations of
the model resulting in the expression for the 1/f noise intensity through
the integral of the characteristic function of the distribution of the time
intervals between the subsequent transition times of the elementary events,
pulses or particles.

It should be noticed, however, that Eq. (25) represent an idealized 1/f
noise. The real systems have finite relaxation time and, therefore,
expression of the noise intensity in the form (23) is valid only for $
f>\left( 2\pi \tau _{rel}\right) ^{-1}$ with $\tau _{rel}$ being the
relaxation time of the period's $\tau _k$ fluctuations. On the other hand,
due to the deviations from the approximation $t_{k+q}-t_k=\tau _kq$ at large
$q$, for sufficiently low frequency we can obtain the finite intensity of $
1/f^\delta $ ($\delta \simeq 1$) noise even in the case $\psi \left(
0\right) =0$ but for the signals with fluctuations resulting in the dense
concentrations of the transit times $t_k$. Generalizations of the approach
(21)--(25) and analysis of the deviations from the idealized 1/f noise
expression (25) are subjects of separate investigations.

\begin{center}
{\bf ACKNOWLEDGMENTS }
\end{center}

Stimulating discussions with Dr. A. Bastys and support from the Alexander
von Humboldt Foundation and Lithuanian State Science and Studies Foundation
are acknowledged.

\vspace{1cm}

\begin{center}
{\bf Captions to the figures of the paper }

\end{center}

\vspace{1cm}

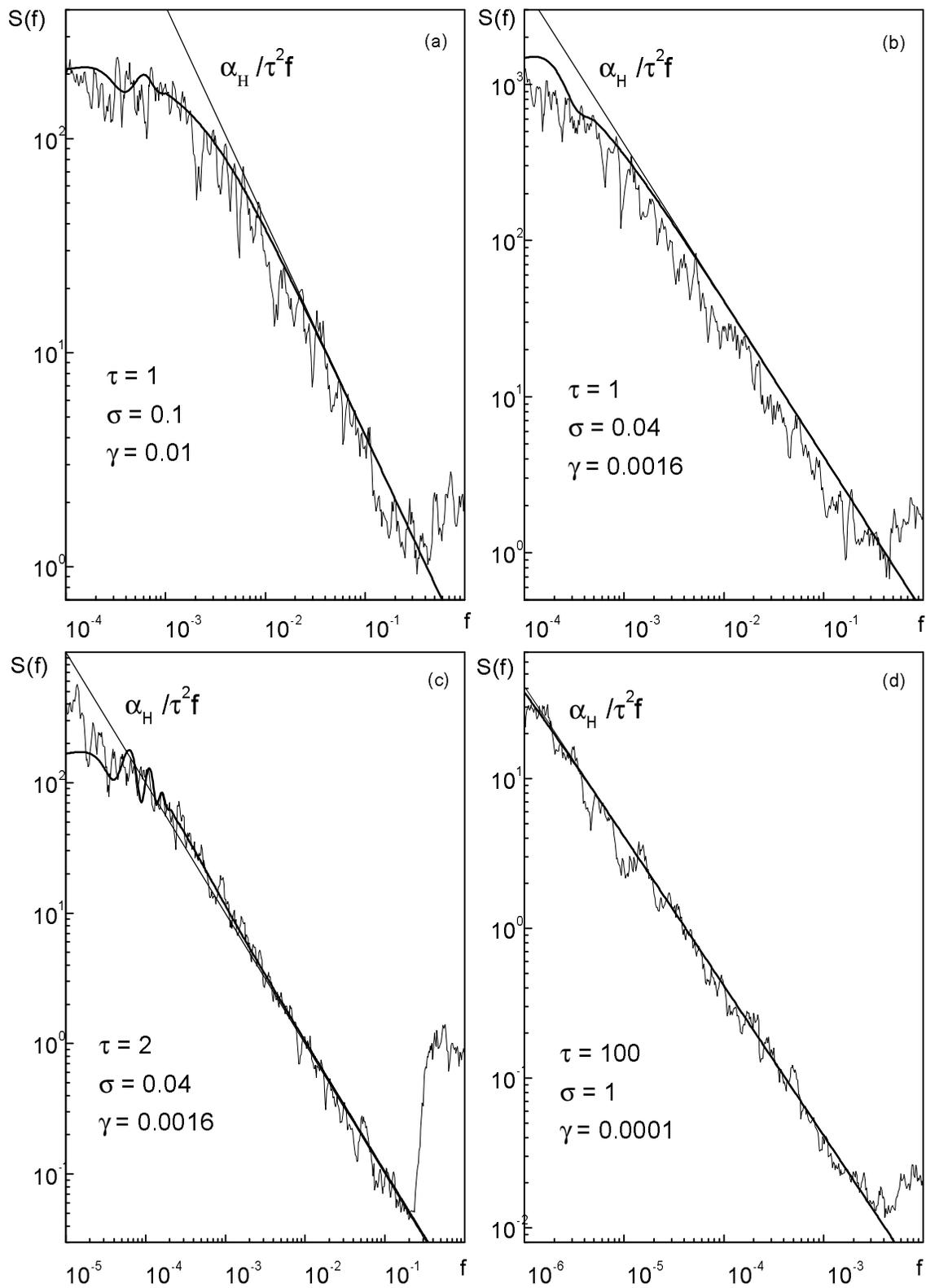
FIG. 1. Power spectral density vs frequency of the current generated by Eqs.
(2)--(4) with the Gaussian distribution of the random increments $\left\{
\varepsilon _k\right\} $ for different parameters $\bar \tau $, $\sigma $,
and $\gamma $. The sinuous fine curves represent the averaged over five
realizations results of numerical simulations, the heavy lines correspond to
the numerical integration of Eq. (16) with $g\left( q\right) $ from Eq.
(13), and the thin straight lines represent the analytical spectra according
to Eqs. (17) and (18).

\vspace{1cm}

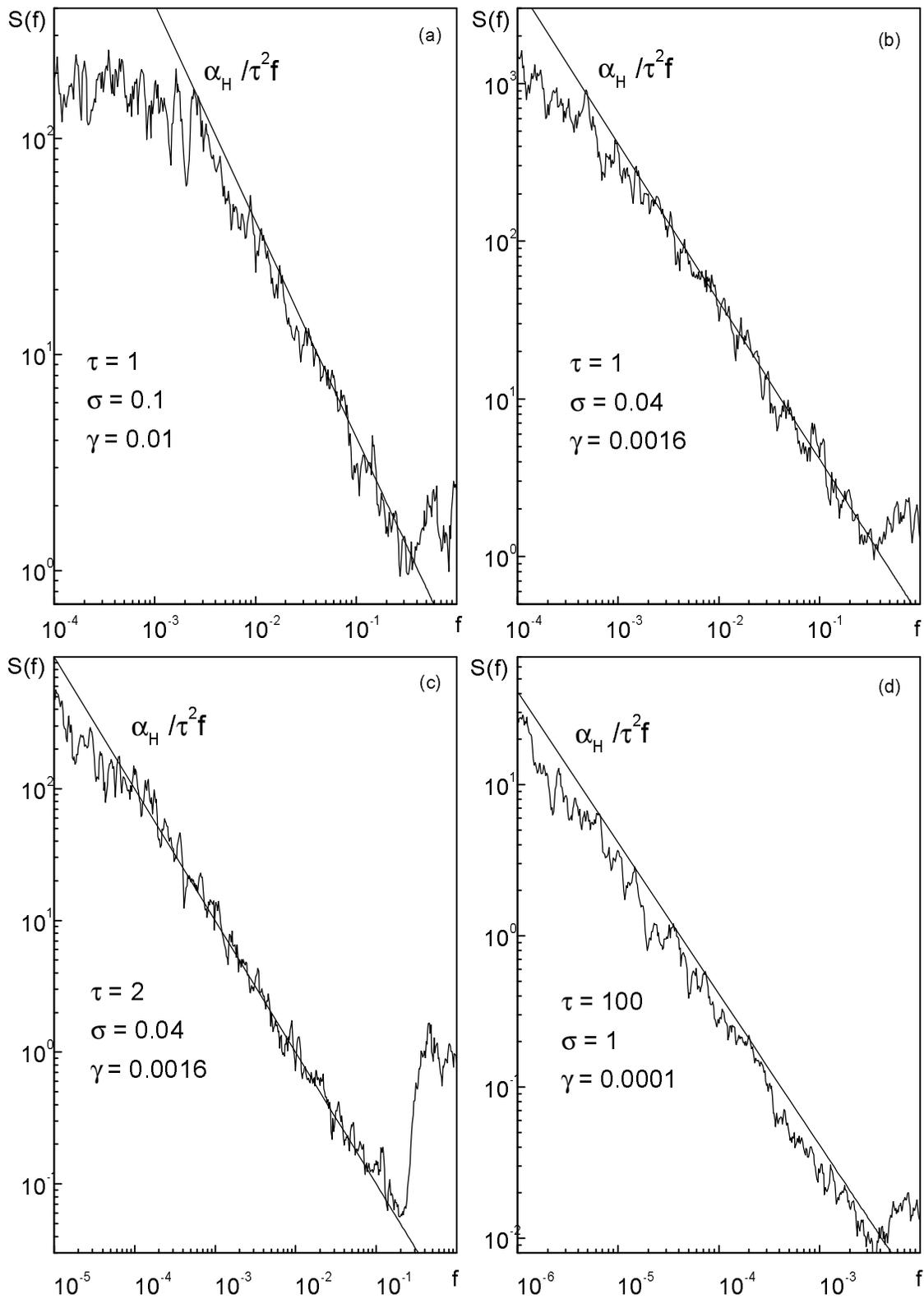
FIG. 2. Same as in Fig. 1 but for the uniform distribution of the random
increments $\left\{ \varepsilon _k\right\} $. Note that for the non-Gaussian
distributions of the random perturbations we have no explicit expression
analogous to Eq. (16) for the integral representation of the noise power
spectral density.

\vspace{1cm}

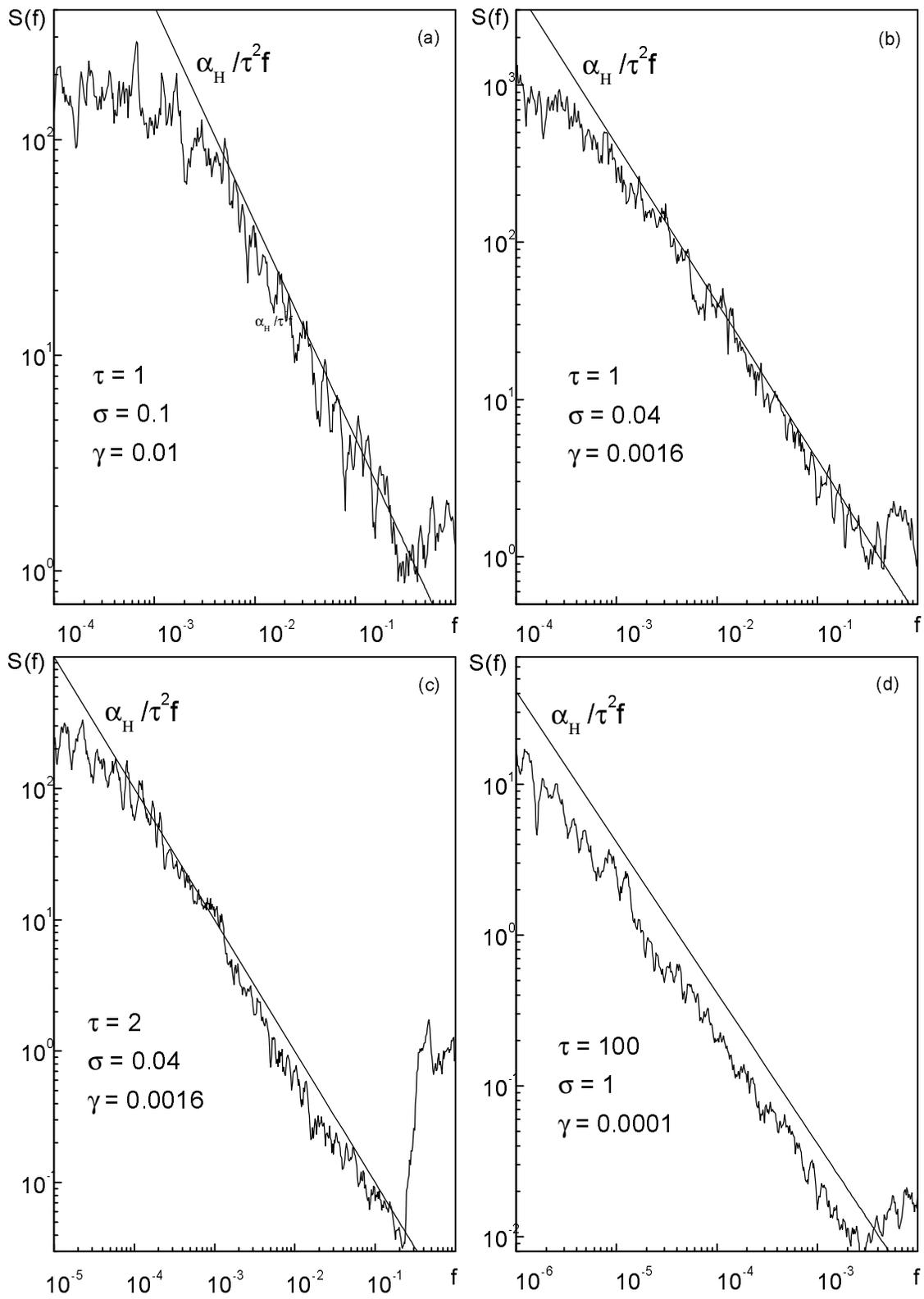
FIG. 3. Same as in Fig. 1 and Fig. 2 but for the asymmetric $\chi _3^2$
distribution of the random increments $\left\{ \varepsilon _k\right\} $.


\begin{references}
\bibitem{hooge81}  F. N. Hooge, T. G. M. Kleinpenning, and L. K. J. Vadamme,
Rep. Prog. Phys. {\bf 44}, 479 (1981); P. Dutta and P. M. Horn, Rev. Mod.
Phys. {\bf 53}, 497 (1981); Sh. M. Kogan, Usp. Fiz. Nauk {\bf 145}, 285
(1985) [Sov. Phys. Usp. {\bf 28}, 170 (1985)]; M. B. Weissman, Rev. Mod.
Phys. {\bf 60}, 537 (1988); M. J. Kirton and M. J. Uren, Adv. Phys. {\bf 38}
, 367 (1989); V. Palenskis, Lit. Fiz. Sb. {\bf 30}, 107 (1990) [Lithuanian
Phys. J. {\bf 30}(2), 1 (1990)]; F. N. Hooge, IEEE Trans. Electron Devices
{\bf 41}, 1926 (1994); G. P. Zhigal'skii, Usp. Fiz. Nauk {\bf 167}, 623
(1997) [Phys.-Usp. {\bf 40}, 599 (1997)].

\bibitem{musha76}  T. Musha and Higuchi, Jap. J. Appl. Phys. {\bf 15}, 1271
(1976); X. Zhang and G. Hu, Phys. Rev. E {\bf 52}, 4664 (1995); M. Y. Choi
and H. Y. Lee, Phys. Rev. E {\bf 52}, 5979, (1995).

\bibitem{press78}  W. H. Press, Comments Astrophys. {\bf 7}, 103 (1978); A.
Lawrence, M. G. Watson, K. A. Pounds, and M. Elvis, Nature (London) {\bf 325}
, 694 (1987); M. Gartner, Sci. Am. {\bf 238}, 16 (1978); S. Maslov, M.
Paczuski, and P. Bak, Phys. Rev. Lett. {\bf 73}, 2162 (1994); M. Usher and
M. Stemmler, ibid {\bf 74}, 326 (1995); Y. Shi, Fractals {\bf 4}, 547
(1996); V. B. Ryabov, A. V. Stepanov, P. V. Usik, D. M. Vavrin, V. V.
Vinogradov, and Yu. F. Yurovsky, Astron. Astrophys. {\bf 324}, 750 (1997).

\bibitem{koba82}  M. Kobayashi and T. Musha, IEEE Trans. Biomed. Eng. {\bf
BME-29}, 456 (1982); R. Voss, Phys. Rev. Lett. {\bf 68}, 3805 (1992); {\bf
76 }, 1978 (1996).

\bibitem{gilden95}  D. L. Gilden, T. Thornton, and M. W. Mallon, Science
{\bf 267}, 1837 (1995).

\bibitem{chen97}  Y. Chen, M. Ding, and J. A. S. Kelso, Phys. Rev. Lett.
{\bf 79}, 4501 (1997).

\bibitem{wolf97}  M. Wolf, Physica A {\bf 241}, 493 (1997).

\bibitem{icnf95}  Proc. 13th Int. Conf. on Noise in Physical Systems and 1/f
Fluctuations, Palanga, Lithuania, 29 May-3 June 1995. Eds. V. Bareikis and
R. Katilius (World Scientific, Singapore, 1995).

\bibitem{icnf97}  Proc. 14th Int. Conf. on Noise in Physical Systems and 1/f
Fluctuations, Leuven, Belgium, 14-18 July 1997. Eds. C. Claeys and E. Simoen
(World Scientific, Singapore, 1997).

\bibitem{upon97}  Proc. 1th Int. Conf. on Unsolved Problems of Noise,
Szeged, Hungary, 3-7 Sept. 1996. Eds. Ch. Doering, L. B. Kiss, and M. F.
Shlesinger (World Scientific, Singapore, 1997).

\bibitem{mandelbr68}  B. B. Mandelbrot and J. W. Van Ness, SIAM Rev. {\bf 10}
, 422 (1968); E. Masry, IEEE Trans. Inform. Theory {\bf 37}, 1173 (1991); B.
Ninness, {\it ibid} {\bf 44}, 32 (1998).

\bibitem{jensen91}  H. J. Jensen, Physica Scripta {\bf 43}, 593 (1991); H.
F. Quyang, Z. Q. Huang, and E. J. Ding, Phys. Rev. E {\bf 50}, 2491 (1994);
E. Milotti, ibid {\bf 51}, 3087 (1995).

\bibitem{kumic94}  J. Kumi\v cak, Ann. Physik {\bf 3}, 207, (1994);
J. Kumi\v cak, in Ref \cite{icnf97}, p. 93; H. Akbane and M. Agu, in
Ref \cite{icnf97}, p. 601 ; J. Timmer and M. K\"onig, Astron. Astrophys.
{\bf 300}, 707 (1995); M. K\"onig and J. Timmer, Astron. Astrophys. Suppl.
Ser. {\bf 124}, 589 (1997).

\bibitem{sinha96}  S. Sinha, Phys. Rev. E {\bf 53}, 4509 (1996); T. Ikeguchi
and K. Aihara, ibid {\bf 55}, 2530 (1997).

\bibitem{kaul97}  B. Kaulakys and G. Vektaris, in Ref \cite{icnf95}, p. 677;
B. Kaulakys and T. Me\v skauskas, in Ref \cite{icnf97}, p. 126;
xxx.lanl.gov/abs/adap-org/9806002.

\bibitem{kaul95}  B. Kaulakys and G. Vektaris, Phys. Rev. E {\bf 52}, 2091
(1995); xxx.lanl.gov/abs/chao-dyn/9504009.

\bibitem{surd39}  J. Bernamont, Ann. Physik {\bf 7}, 71 (1937); M. Surdin,
J. Phys. Radium (Serie 7) {\bf 10}, 188 (1939); F. K. du Pr\'e, Phys. Rev.
{\bf 78}, 615 (1950); A. Van der Ziel, Physica {\bf 16}, 359 (1950); A. L.
McWhorter, in: {\it Semiconductor surface physics,} edited by R. H. Kingston
(Univ. Penn., Philadelphia, 1957), p. 207; F. N. Hooge, in Ref \cite{icnf97}
, p. 3.

\bibitem{lukes61}  T. Lukes, Proc. Phys. Soc. (London) {\bf 78}, 153 (1961);
C. Heiden, Phys. Rev. {\bf 188}, 319 (1969); K. L. Schick and A. A. Verveen,
Nature (London) {\bf 251}, 599 (1974).

\bibitem{kaul98}  B. Kaulakys, xxx.lanl.gov/abs/adap-org/9806004 and to be
published; B. Kaulakys and T. Me\v skauskas, 7th Vilnius Conf. on Probab.
Theory and 22nd European Meeting of Statisticians, Vilnius, August 12-18,
1998. Abstracts (Vilnius, TEV, 1998), p.265; A. Bastys and B. Kaulakys,
ibid, p. 144.
\end{references}
\end{document}